\documentclass[copyright]{eptcs}
\providecommand{\event}{ACAC 2009} 
\providecommand{\volume}{??} \providecommand{\anno}{20??}
\providecommand{\firstpage}{1} \providecommand{\eid}{??}
\usepackage{breakurl}        
\usepackage{latexsym} \usepackage{amssymb,amstext,amsmath}
\usepackage{color}

\def\Proof{{\noindent \bf Proof. }}
\def\QED{{\hfill $\Box$ \medskip}}

\newtheorem{Def}{Definition}
\newtheorem{Lem}{Lemma}
\newtheorem{Prop}{Proposition}
\newtheorem{Theo}{Theorem}
\newtheorem{Cor}{Corollary}

\usepackage{algorithmic}
\usepackage{algorithm}
\usepackage{graphicx}
\def\NN{\hbox{\sf I\kern-.13em\hbox{N}}}

\renewcommand{\algorithmicfor}{\textbf{For}}
\renewcommand{\algorithmicendfor}{\textbf{EndFor}}
\renewcommand{\algorithmicensure}{\textbf{Return:}}
\newcommand{\mc}[1]{\mathcal{#1}}
\newcommand{\chg}[1]{{#1}}
\newcommand{\suppr}[1]{}

\title{Cartesian product of hypergraphs: properties and algorithms}
\author{Alain  Bretto \email{alain.bretto@info.unicaen.fr} \and Yannick Silvestre \email{yannick.silvestre@info.unicaen.fr}  \and Thierry Vall\'ee \email{vallee@pps.jussieu.fr}
\institute{Universit\'e de Caen, GREYC CNRS UMR-6072, Campus
II, Bd Marechal Juin BP 5186, 14032 Caen cedex, France.}}
\def\titlerunning{Cartesian product of hypergraphs: properties and algorithms}
\def\authorrunning{A. Bretto \& Y. Silvestre \& T. Vall\'ee}
\begin{document}
\nocite{*}
\maketitle

\begin{abstract}
Cartesian products of graphs have been studied extensively since the 1960s. They make it possible to decrease the algorithmic complexity of problems by using the factorization of the product. Hypergraphs were introduced as a generalization of graphs and the definition of Cartesian products extends naturally to them. In this paper, we give new properties and algorithms concerning coloring aspects of Cartesian products of hypergraphs. We also extend a classical prime factorization algorithm initially designed for graphs to connected conformal hypergraphs using 2-sections of hypergraphs.
\end{abstract}

\section{Introduction}
Cartesian products of graphs have been studied since the 1960s by Vizing and Sabidussi. In \cite{Vizing} and \cite{Sabi} they independently showed, among other things, that for every finite connected graph there is a unique (up to isomorphism) prime decomposition of the graph into factors. This fundamental theorem was the starting point for research concerning the relations between a Cartesian product and its factors \cite{Zerovnik,ImPiZe,AC,Lauri}.
 Some of the questions raised are still open, as in the case of the Vizing's conjecture\footnote{This conjecture expressed by Vizing in 1968 states that the domination number of the Cartesian product of graphs is greater than the product of the domination numbers of its factors.}.
These relations are of particular interest as they allow us to break down problems by transferring algorithmic complexity from the product to the factors. In 2006, Imrich and Peterin \cite{ImPe} gave an algorithm able to compute the prime factorization of connected graphs in linear time and space, making the use of Cartesian product decomposition particularly attractive.

Most of networks used in the context of parallel and distributed computation are Cartesian products: the hypercube, grid graphs, etc. In this context, the problem of finding a ``Cartesian'' embedding of an interconnection
network into another is also of fundamental importance and thus has gained considerable attention. Cartesian products are also used in telecommunication \cite{Vesel}.\smallskip

Hypergraph theory has been introduced in the 1960s as a generalization of graph theory. A lot of applications of hypergraphs have been developed since (for a survey see \cite{Bretto}). Cartesian products of hypergraphs can be defined in a same way as graphs, and similarly it is easier to study the hypergraph factors than the product.

In this paper, we give some new properties and algorithms concerning coloring aspects of Cartesian products of hypergraphs.
We show that the algorithm of Imrich and Peterin \cite{ImPe} can be used to find the prime factorization of connected conformal hypergraphs by considering the \emph{2-section} and the \emph{labelled 2-section} of the hypergraph to be factorized.\\

\section{Preliminaries}
The general terminology concerning graphs and hypergraphs in this
article is similar to the one used in \cite{berge1,berge2}. The
cardinality of a set $A$ is denoted by $|A|$. For $f$ a function,
we define $\textrm{Im} (f) = \{ y: \exists x, f(x)= y\}$.

A \emph{hypergraph} $H$ on a set of vertices $V$ is a pair $(V,E)$
where $E$ is a set of non-empty subsets of $V$ called
\emph{hyperedges} such that $\bigcup E =V$. This implies in
particular that every vertex is included in at least one
hyperedge. A hypergraph is \emph{simple} if no hyperedge is
contained in another. In the sequel, we suppose that hypergraphs
are simple and that no edge is a loop, that is, the cardinality of
a hyperedge is at least $2$. The number of hyperedges of a
hypergraph $H$ is denoted by $m(H)$. We sometime write $\mc V(H)$
to denote the set of vertices of a hypergraph and $\mc E(H)$ for
its set of edges. Given a hyperedge $e \in \mc E(H)$, we sometime
write $\mc P_2(e)$ to denote the pairs of vertices of $e$. A
\emph{graph} is a particular case of simple hypergraph where every
(hyper)edge is of size 2.

\chg{Given a graph $\Gamma =(V,E)$ and $A$ a subset of $E$, we
define $\Gamma (A) = (S,A)$ as a subgraph of $\Gamma$ where $S =
\{x \in  a \in A\}$. }

The \emph{ degree of} $x$ (denoted by $d(x)$) is $|H(x)|$ and
$\Delta(H)$ is the maximal degree of a vertex in $H$.
A \emph{k-coloring} of a hypergraph $H$ is \chg{a map $f : \mc V(H) \rightarrow \mathbb{N}$ such that $|\textrm{Im}(f)|=k$ and  
such that every hyperedge $e \in \mc E(H)$ 
has two vertices $x, y \in e$, with $f(x) \neq f(y)$.} The
\emph{chromatic number}, denoted by $\chi(H)$, is the smallest
integer $k$ for which $H$ admits a $k$-coloring. A \emph{strong
k-coloring} of $H$ is \chg{{a map $f : \mc V(H) \rightarrow
\mathbb{N}$ such that $|\textrm{Im}(f)|=k$ and such that every
hyperedge $e \in \mc E(H)$ verifies: $\forall x, y \in e$, $f(x)
\neq f(y)$.}}

Let $H$ be a hypergraph, the \emph{chromatic index} of $H$ is the least number of colors necessary to color the hyperedges of $H$ such that two intersecting hyperedges are always colored differently. This number is denoted by $q(H)$. It is easy to see that $q(H) \geq \Delta(H)$. A hypergraph $H$ has the \emph{colored hyperedge property} if  $q(H) = \Delta(H)$.

For $E'\subseteq E$ the set $H'= (\bigcup E' , E')$ is the \emph{partial hypergraph} generated by $E'$. 

The \emph{2-section} $[H]_2$ of a hypergraph $H$ is the graph
whose vertices are the vertices of $H$, and where two vertices are
adjacent iff they belong to a same hyperedge. Notice that every
hyperedge of $H$ is a clique of $[H]_2$. A hypergraph $H$ is
\emph{conformal} if, for every $e\subseteq \mc V(H)$, $e$ is a
maximal clique of $[H]_2$ iff $e$ is a hyperedge of $H$.

Finally, an isomorphism from the hypergraph $H=(V,E)$ to the
hypergraph $H'=(V',E')$ is a bijection from $V$ to $V'$ such that,
for every $e \subseteq V$, $e \in E$ iff $\{f(x) : x \in e\} \in
E'$. Note that the isomorphism $f$ induces a bijection $f^{\#} : E
\rightarrow E'$ defined by $f^{\#}(e) = \{f(x) : x \in e\}$, for
every $e\in E$.

\section{Cartesian Product of  Hypergraphs: definition and coloring properties.}

Let $H_1=(V_1, E_1)$ and  $H_2=(V_2, E_2)$
be hypergraphs. The \emph{Cartesian product} of $H_1$ and $H_2$ is the hypergraph $H_1 \Box H_2$ with set of vertices
$V_1 \times V_2$ and set of edges:
\[
 E_1\Box E_2= \underbrace{\{ \{x\}\times e : x\in V_1 \;\mbox{and}\;  e\in E_2\}}_{A_{1}} \cup\underbrace{ \{e\times
 \{u\}: e\in E_1 \; \mbox{and}\;  u\in V_2\}}_{A_{2}}.
\]

Note that up to the isomorphism the Cartesian product is
commutative and associative. That allows us to denote simply by
$v_1 \ldots v_k$ the vertices of $V_1 \times \ldots \times V_k$.
\suppr{SE REPETE AVEC LA PREUVE DU LEMME Note also that a
hyperedge $\{x\} \times e$ of $A_1$ and a hyperedge $e' \times
\{u\}$ of $A_2$ have a non-empty intersection iff $x \in e'$ and
$u\in e$. In that case $\{x\} \times e \cap e' \times \{u\}=
\{xu\}$.}

In the sequel $H=(V,E)$ will always stand for the Cartesian product of two hypergraphs $H_1 = (V_1,E_1)$ and $H_2 = (V_2,E_2)$, unless explicitely stated. We use $x,y,z$ to denote the vertices of $V_1$ and $u,v,w$ to denote the vertices of $V_2$.

\begin{Lem}\label{Fond}
$A_{1}\cap A_{2} = \emptyset$. Moreover, $\vert e\cap e'\vert \leq 1$ for any $e\in A_{1}$ and any $e' \in A_2$.
\end{Lem}

\begin{Prop}\label{Prop2Section}

The 2-section of $H$ is the Cartesian product of the 2-section of $H_1$ and the 2-section of $H_2$
\end{Prop}

\begin{Theo}\label{ThmConformalCartesianStable} If $H = H_1\Box H_2$ then $H$ is conformal if and only if  $H_1$ and  $H_2$ are conformal.
\end{Theo}

\noindent We give now two new results about coloring aspects of Cartesian products of hypergraphs.

\begin{Theo}

 If $H_1$ and $H_2$ \chg{have} both the colored hyperedge property then $H$ has the colored hyperedge property.
\end{Theo}

\begin{Theo} Let $\chi _i$ and $\chi$ (respectively $\gamma _i$ and $\gamma$) be the chromatic number (resp. the strong chromatic number) of $H_i$ and $H$. We have the following:
\begin{enumerate}
\item $\chi =\max\{\chi _1; \chi _2\}$
\item $\gamma =\max\{\gamma _1; \gamma _2\}$.
\end{enumerate}
\end{Theo}

This leads to a straightforward algorithm to compute a minimal coloring (Algorithm \ref{COL}) of a given hypergraph $H$ thanks to the minimal colorings of its factors. In the case the hypergraph is prime, there is no other choice but to use classical coloring algorithms. As the problem of determining if the chromatic number of a given hypergraph is less than an integer $k$ is yet NP-complete (for $k \geqslant 2$), it is worthwhile to add this preliminary step at the beginning of the investigation. An algorithm specifying how to compute the factors for conformal hypergraphs will be given in the following pages.

    \begin{algorithm}
        \caption{Coloring algorithm COL}
        \label{COL}
        \begin{algorithmic}[1]
        \REQUIRE A hypergraph $H=(V,E)$.
        \ENSURE A $H$-coloring $f$.
    \STATE Find a factorization in prime hypergraphs $H = H_1 \Box H_2 \Box \ldots \Box H_k$.
    \STATE Compute for each prime factor $H_i$ a minimal coloring
    $f_i$ with $|\textrm{Im}(f_i)| = \chi _i$.
    \RETURN $f$ such that $f(x_1, \ldots, x_k)=  \sum _{i=1}^k f_{i}(x_i) \mod \max _{i=1}^k \chi _i$
    \end{algorithmic}
    \end{algorithm}

\section{Hypergraph factorization algorithm}
In the sequel, we use some of the results from Imrich and Peterin in \cite{ImPe}. In order to find prime factorizations of conformal hypergraphs, we extend the algorithm given in \cite{ImPe}. This algorithm is based on a coloring of the edges of the graph $G$ to be factorized in such a way that the factors are proper subgraphs of $G$ said \emph{layers}. It uses the fact that if $G = G_1 \square \dotsc \square G_k$ is a Cartesian product of graphs then, for all $w \in V_1 \times \dotsc \times V_k$ and $i \in \{1, \dotsc ,k\}$, there is a subgraph $G^w_i$ of $G$ such that the $i^{th}$ projection $p_i$ induces an isomorphism between $G^w_i$ and $G_i$. Indeed, we remark that $\{w,w'\}$ is an edge of $G$ iff there exists some $i \in \{1, \dotsc ,k\}$ such that $\{p_i(w), p_i(w')\}$ is an edge of $G_i$ and $w,w'$ differ only on their $i^{th}$ coordinates. The graph $G^w_i$ is then defined as the graph whose vertices are the $k$-tuples which differ from $w$ at most on the $i$-th coordinate, and where $\{w',w''\}$ is an edge of $G_i^w$ iff $\{p_i(w), p_i(w')\}$ is an edge of $G_i$.

Subgraphs of the form $G^w_i$ are the layers of $G$ and it can be easily shown that every edge of $G$ is contained in exactly one layer. Moreover the edge sets of the layers partition the edge set of $G$.

\chg{We recall the square lemma, given in \cite{ImPe}:}

\begin{Lem}[Square lemma]\label{LmSq}
 \chg{If two edges are adjacent edges which belong to non-isomorphic layers, then these edges lay in a unique induced square.}
\end{Lem}

 A \chg{straightforward} consequence of the Square Lemma given in \cite{ImPe} is that every triangle of $G$ is necessarily contained in the same layer. From these facts we get easily the following result.

\begin{Lem}\label{LmCliqueInLayer} Let $G = G_1 \square \dotsc \square \chg{G_k}$ be the Cartesian product of graphs. Then every clique of $G$ is contained in the same layer. Moreover, if two cliques share an edge then they are both contained in the same layer.
\end{Lem}

The extension to hypergraphs of the algorithm of \cite{ImPe} uses L2-sections. We start by some general definitions and basic properties then we extend Cartesian products and isomorphisms to L2-sections.

\begin{Def}\label{DfL2Section} Let $H=(V,E)$ be a hypergraph, we define the $L2$-section $[H]_{L2}$ of $H$ as the triple $\Gamma =(V,E',\mc L)$ where  $(V,E')$ is the 2-section of $H$ and $\mc L : E' \rightarrow \mc P (E)$ is defined by $\mc L(\{x,y\}) = \{e: x,y \in e\in E\}$.\\
\end{Def}

Hence, the $L2$-section of a hypergraph is a labelled version of the 2-section where every edge $\{x,y\}$ is labelled with the set of hyperedges containing $x$ and $y$. In that way, it is possible to keep track of all the hyperedges from which the edge $\{x,y\}$ comes from. It is then possible to build back the hypergraph from its labelled 2-section as shown in the definition below.

\begin{Def}\label{DfL2SectionToHypergraph} Let $\Gamma = (V,E', \mc L)$ be a $L2$-section, we define the hypergraph $[\Gamma]^{-1}_{L2} = (V,E)$ by $E= \bigcup
\textrm{Im}(\mc L)$(see definition of $\textrm{Im}$ in section
Preliminaries).
\end{Def}

\noindent Not surprisingly, from the two definitions above, we get easily:

\begin{Prop}\label{PrInjectivityL2} For all hypergraphs $H$ and L2-sections $\Gamma$ we have
$[[H]_{L2}]_{L2}^{-1}=H$ and $[[\Gamma]^{-1}_{L2}]_{L2}= \Gamma$.
\end{Prop}

We extend now the Cartesian product operation to L2-sections.

\begin{Def}\label{DfProductLabeledGraph} Let $\Gamma_1=(V_1,E_1',\mc L_1)$ and $\Gamma_2= (V_2,E_2',\mc L_2)$ be the $L2$-sections of $H_1= (V_1, E_1)$ and $H_2=(V_2,E_2)$. We define their Cartesian product $\Gamma_1 \square \Gamma_2$ as the triple $(V, E', \mc L_1 \Box \mc L_2)$ where:
\begin{itemize}
 \item $(V,E')$ is the Cartesian product of $(V_1,E_1')$ and $(V_2,E_2')$.
 \item $\mc L_1 \Box \mc L_2$ is the map from $E = E_1' \Box E_2'$ to $\mc P (E_1 \Box E_2))$ defined by:
\[ \mc L_1 \Box \mc L_2 (\{(x,u),(y,v)\})= \left\{ \; \begin{array}{ll} \{\,\{x\} \times e : e \in \mc L_2(\{u,v\}) \, \} & \textrm{if $x = y$.}\\
                                        \{\,e \times \{u\} : e \in \mc L_1(\{x,y\})\,\} & \textrm{if $u = v$.} \end{array} \right. \]
\end{itemize}
\end{Def}

Note that the definition of $\mc L_1 \Box \mc L_2$ above is
correct. Indeed, by definition of the L2-section, for every edge
$\{(x,u),(y,v)\}$ of $E_1' \Box E_2'$ there exists an hyperedge
$\varepsilon \in E_1 \Box E_2$ such that $(x,u),(y,v) \in
\varepsilon$. Now, by definition of $H_1 \Box H_2$, either
$\varepsilon = \{x\} \times e$, where $x \in V_1$ and $e \in E_2$,
or $\varepsilon = e \times \{u\}$, where $u \in V_2$ and $e \in
E_1$. In the first case we have then $x =y$ and so $u\neq v$
(otherwise $\{(x,u),(y,v)\}$ would be a loop), and in the second
$u=v$ and so $x \neq y$. It is moreover easy to check that $\mc
L_1 \Box \mc L_2(\{(x,u),(y,v)\}) \subseteq E_1 \Box E_2$.

\begin{Lem}\label{LemSCommuteWithCartesian} For all hypergraphs $H_1, H_2$ we have:
\begin{enumerate}
\item $[H_1 \Box H_2]_{L2} = [H_1]_{L2} \Box [H_2]_{L2}$\\[2pt]
\item $[[H_1]_{L2} \Box [H_2]_{L2}]_{L2}^{-1} = [[H_1]_{L2}]_{L2}^{-1} \Box [[H_2]_{L2}]^{-1}_{L2}$.
\end{enumerate}
\end{Lem}

\begin{Def}\label{DfSubsection} Let $\Gamma= (V,E',\mc L)$ be the $L2$-section of $H$. A triple $\Gamma_0 = (V_0, E'_0, \mc L_0)$ is a \emph{subsection} of $\Gamma$ if the following conditions are satisfied:
\begin{enumerate}
\item 
\chg{$E'_0$ is a subset of $E'$ and $(V_0,E'_0) = \Gamma (E_0')$.}
\item $\mc L_0$ is the restriction of $\mc L$ to $E'_0$. \item
\chg{If $e \in \bigcup \textrm{Im}(\mc L_0)$, then $\mc P_2(e) \subset
E'_0$.}
\end{enumerate}
\end{Def}

It is easy to check that if $\Gamma_0$ is a subsection then it is
the L2-section of the hypergraph $H_0 = [\Gamma_0]^{-1}_{L2}$. It
is also easy to verify that $H_0$ is a partial hypergraph of $H$.
We have moreover the following.

\begin{Lem}\label{LmSubsectionConformal} Let $\Gamma$ be the L2-section of the conformal hypergraph $H$ and $\Gamma_0$ be a subsection of $\Gamma$. Then $H_0 = [\Gamma_0]^{-1}_{L2}$ is a conformal partial hypergraph of $H$.
\end{Lem}

\begin{Def}\label{DfIsomorphismL2Section} An isomorphism between two $L2$-sections $\chg{\Gamma_1} = (\chg{V_1}, \chg{E_1}, \chg{\mc L_1})$ and $\chg{\Gamma_2}=(\chg{V_2}, \chg{E_2}, \chg{\mc L_2})$ is a bijection $f$ from \chg{$V_1$} to \chg{$V_2$} such that:
\begin{itemize}
 \item $\{x,y\} \in \chg{E_1}$ if and only if $\{f(x),f(y)\} \in \chg{E_2}$, for all $x,y \in \chg{V_1}$.
 \item $e \in \chg{\mc L_1}(\{x,y\})$ if and only if $\{f(z) : z \in e \} \in \chg{\mc L_2}(\{f(x),f(y)\})$, for all $x,y \in \chg{V_1}$ and $e \subseteq \chg{V_1}$.
\end{itemize}
We write $\chg{\Gamma_1} \cong \chg{\Gamma_2}$ to express that there exists an isomorphism between $\chg{\Gamma_1}$ and $\chg{\Gamma_2}$.
\end{Def}

\begin{Lem}\label{LmIsoEquivSection} Let $H$ and $H'$ be two hypergraphs. The two first statements below are equivalent. If moreover $H$ and $H'$ are conformal then they are equivalent to the third one.
\begin{enumerate}
 \item $H \cong H'$
 \item $[H]_{L2} \cong [H']_{L2}$
 \item $[H]_{2} \cong [H']_{2}$
\end{enumerate}
\end{Lem}

By combining the first point of Lemma
\ref{LemSCommuteWithCartesian} and the two first points of Lemma
\ref{LmIsoEquivSection}, it is straightforward to check that, up
to isomorphism, the Cartesian product is commutative and
associative on L2-sections. That allows us to overlook parenthesis
for Cartesian products of L2-sections. We give now a last
essential lemma before we introduce the factorization algorithm
for conformal hypergraphs.

\begin{Lem}\label{LmPassageAuProduitEtiquetage} Let $\Gamma = (V,E',\mc L)$ be the L2-section of the conformal hypergraph $H$ and let $G$ be its 2-section. Suppose $G = G_1 \Box G_2$ where \chg{$G_i=(V_i,E_i')$, $i \in \{1,2\}$} are \chg{layers in} $G$. Define  $\Gamma _ i = (V_i, E'_i , \mc L_i)$, where $\mc L_i$ is the restriction of $\mc L$ to $E_i'$. Then $H_i=[\Gamma _i ]^{-1}_{L2}$ is a conformal partial hypergraph of $H$. Moreover we have $H = H_1 \Box H_2$.
\end{Lem}

We introduce now an algorithm which gives the prime factorization of conformal hypergraphs. The idea is the following. From the connected hypergraph $H$ it first builds the $L2$-section $\Gamma$ of $H$. Then it runs the algorithm of Imrich and Peterin which colors the edges of the unlabelled underlying graph $G$ with color $i$ all edges of all layers that belong to the same factor $H_i$. When obtained the factorization $G_1, \dotsc ,G_k$ of $G$, the labels of the edges of the $G_i$'s are used to build hypergraphs $H_1, \dotsc , H_k$ they come from.

    \begin{algorithm}[!h]
        \caption{Hypergraph-prime decomposition}
        \label{algo2}
        \begin{algorithmic}[1]
        \REQUIRE A conformal hypergraph $H=(V,E)$.
        \ENSURE The prime factors of $H$, that is $H_1,H_2, \ldots, H_k$ such that
        $H = H_1 \Box H_2 \Box \ldots \Box H_k$.
    \STATE Compute $\Gamma =(V,E',\mc L)$, the L2-section of $H$.
    \STATE Run the algorithm of prime-decomposition on the underlying graph $G=(V,E')$ and call $G_1 =(V_1,E_1'), \ldots, G_k=(V_k, E_k')$ its prime factors such that $G = G_1 \Box G_2 \Box \ldots \Box G_k$.
    \STATE Define $\mc L_i$ as the restriction of $\mc L$ to $E_i'$, $\Gamma_i = (V_i,E_i',\mc L_i)$, and build $H_i = [\Gamma_i]^{-1}_{L2}$, for every $i \in \{1, \dotsc, k\}$.
        \RETURN $H_1,H_2, \ldots, H_k$
    \end{algorithmic}
    \end{algorithm}

\begin{Theo} Algorithm 2 is sound and complete for every conformal hypergraph $H$.
\end{Theo}

As the algorithm of Imrich and Peterin is able to return the factors of connected graphs with respect to the Cartesian product in linear time and space, the overall complexity of the given algorithm for a hypergraph $H$ is in $O(m(H)r(H)^2)$, as 2-section operations demand up to $r(H)^2$ per hyperedge.

\chg{
\begin{Cor}[Corollary of lemma \ref{LmIsoEquivSection}]
 Given a conformal hypergraph $H$, there is a unique decomposition in prime factors with respect to the Cartesian product (up to isomorphisms).
\end{Cor}
\Proof
 The uniqueness of the decomposition of the 2-sections, imply the uniqueness of the decompositionn of the conformal hypergraphs they come from. \QED}

\section{Conclusion: Cartesian product and algorithmic complexity of problems}
In the previous section we dealt with some properties which were Cartesian product stable. The question we are interested in here is whether, in order to solve a decision problem $\mathcal P$, it is possible to design an operator on the hypergraph space which fulfill the following conditions:
\begin{itemize}
 \item The operator $\phi$ must connect a plain hypergraph to a Cartesian product.
 \item For any hypergraph $H$, $\phi$ must introduce a relation between $\phi (H)$ and $H$ about $\mc P$, and this relation must be polynomially evaluable from $\phi (H)$ (that is to say, if one knows $\phi (H) \in \mc P$ or $\phi (H) \notin \mc P$, one has to be able to compute in polynomial time whether $H$ is in $\mc P$ or not). In the ideal case, this relation is constant-time or linear-time evaluable.
 \item The operator $\phi$ has to be designable in polynomial time (relatively to the size or the order of the hypergraph).
\end{itemize}
Designing such operators is an interesting issue. Once such operators are built, assuming that no polynomial algorithm is known to solve the problem $\mc P$, it is possible to design competitive algorithms running on the hypergraph factors rather than on the whole Cartesian product.\\

\bibliographystyle{alpha}
\bibliography{bib}

\end{document}